\documentclass[aps,pre,twocolumn,superscriptaddress]{revtex4}
\usepackage[utf8]{inputenc}
\usepackage{amsmath}
\usepackage{amssymb}
\usepackage{mathtools}
\usepackage{graphicx}
\usepackage{xcolor}
\usepackage{physics}

\usepackage{hyperref}

\usepackage{hyperref}

\begin{document}

\title{Minigap suppression in S(N/F)S junctions}

\author{P.\ A.\ Ioselevich}
\affiliation{National Research University Higher School of Economics, 101000 Moscow, Russia}
\affiliation{L.\ D.\ Landau Institute for Theoretical Physics, Kosygin str.\ 2, Moscow, 119334 Russia}

\author{D.\ A.\ Chuklanov}
\affiliation{National Research University Higher School of Economics, 101000 Moscow, Russia}

\begin{abstract}
We consider a long diffusive Josephson junction where the weak link is a thin normal metal (N) - ferromagnetic (F) bilayer (N and F form parallel links between the superconductors (S)). We show that superconductivity in the weak link can be described by an effective one-dimensional Usadel equation containing a "diluted" exchange field as well as a weak depairing term that is caused by the inherent inhomogeneity of the bilayer. The depairing mechanism distinguishes the S(N/F)S system from an SFS junction and affects the density of states of the S(N/F)S junction. It results in the suppression of the minigap in the spin-resolved density of states. The depairing rate and the minigap are expressed in terms of geometrical parameters, the Thouless energy and the effective exchange field.  The effective one-dimensional theory can be applied to various structures with thin inhomogenous links and shows good agreement with numerical solutions of the original two-dimensional equations. We also discuss ways to reveal the predicted effect experimentally. 
\end{abstract}

\maketitle

\begin{figure}
\centering
\hspace*{-3pt}\includegraphics[width=0.485\textwidth]{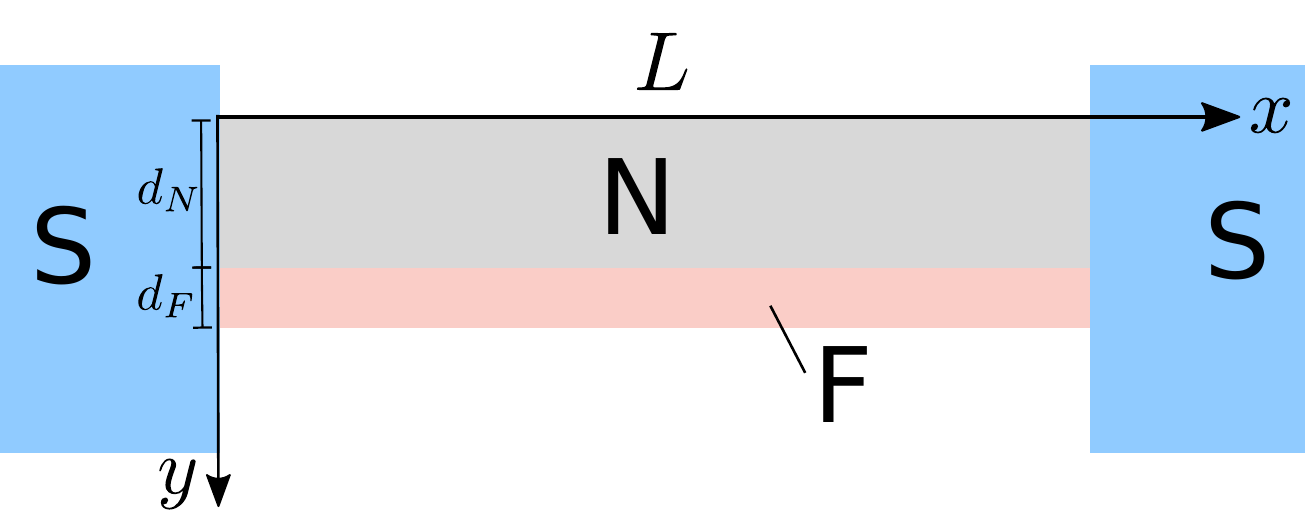}
\caption{S(N/F)S system. The N and F layers are thin in the $x$ direction compared to junction length in the $y$ direction: $d_N,d_F\ll L$. All interfaces are transparent. }
\label{figSystem}
\end{figure}
\textit{1.\ Introduction ---} Heterostructures involving superconductors and ferromagnets have long been studied theoretically and experimentally. The interplay of the two competing orders leads to many interesting phenomena (for reviews see \cite{BuzdinReview,MironovReview,BergeretReview}). The proximity effect in SF structures exhibits oscillating behavior: the induced anomalous pairing in F not only decays away from S but also alternates sign on the same length scale $l_h$. Thus, in SFS geometry, depending on ferromagnetic link length $L$ vs $l_h$ the Josephson junction may be in the $\pi$-contact regime \cite{pi1,pi2}, where the ground state corresponds to $\pi$ phase difference. Using multiple ferromagnets of different polarization gives rise to even more phenomena. For instance, FFS or FSF sandwiches can act as superconducting spin valves \cite{SSV1,SSV2,SSV3}. 

All these applications share a technological problem: the exchange field $h$ in most ferromagnets is relatively high and the decay length $l_h=\sqrt{D/h}$ (where $D$ is the diffusion constant) is rather short. Thus, to get consistent, predictable results, samples have to be prepared very precisely since an error of $\sim l_h$ can drastically change the behaviour of the system. One possible workaround is to effectively dilute the ferromagnet with normal metal. If the ferromagnetic link is replaced with an NF bilayer, as shown on Fig. \ref{figSystem}, Cooper pairs diffusing through the link will spend part of the time in F and part of the time in N. The exchange field $h$ only affects the Cooper pair in F. Thus, on average a Cooper pair in the bilayer link experiences an effective field $h_\mathrm{eff}$ that is smaller than $h$. \cite{Golikova2012}. 

In this paper we construct a theory for diffusive S(N/F)S junctions where the weak link is an NF bilayer as shown on Fig.~\ref{figSystem}. The layers are assumed thin, which allows us to derive an effective one-dimensional equation describing superconductivity in the system. By iteratively including corrections stemming from transverse inhomogeneity we derive all the relevant terms in the effective equation. In particular, we confirm that the one-dimensional equation includes an effective diluted exchange field $h_\mathrm{eff}$. Moreover, we also find a depairing term (similar to terms representing spin-flip events) in the next order in layer thickness. This depairing term manifests itself in the density of states of the S(N/F)S junction, distinguishing it from a regular SFS junction.

Systems of geometry Fig.~\ref{figSystem} or similar have been previously studied theoretically in a variety of cases described by a linearized Usadel equation \cite{Usadel}\cite{Karminskaya2007,Karminskaya2009,Karminskaya2010}. Ref.~\cite{Karminskaya2012} considered partially non-linear equations in a short junction. We consider long junctions and fully non-linear Usadel equations. In particular, we will study properties of the minigap \cite{minigapGolubov,minigapAltland,minigapDevoret} which is a phenomenon dependent on non-linearity. 

We begin by introducing the general framework and parametric regime we work in. We then show how the F layer can be reduced to an effective boundary condition (b.c.) for the equation in N. We then derive the effective one-dimensional Usadel equation. We explain how this depairing affects the density of states in the system and compare our theoretical results with numerics and existing experiment. Lastly we discuss the applicability and scope of our effective theory beyond S(N/F)S junctions and present our conclusion.

\textit{2.\ Model --- } We consider a long junction with a diffusive weak link consisting of a normal (N) and a ferromagnetic (F) layer, as shown on Fig. 1. The length of the junction is $L$ and the layer thicknesses are $d_N,d_F$. The system is assumed to be in the dirty limit, where it can be described by the Usadel equation \cite{Usadel}. We assume the junction to be long so that $L\gg d_N,d_F,\sqrt{D_N/\Delta}$ where $D_N$ is the diffusion constant in N and $\Delta$ is the order parameter in S. The ferromagnetic is single-domain so that $\mathbf{h}=\mathrm{const}$ in F and spin projection $\sigma$ along $\mathbf{h}$ direction is conserved. In this case electrons with different $\sigma$ do not mix and are described by separate equations. For simplicity we focus on the case of zero superconducting phase difference. In this case the Green's function for each spin $\sigma$ is parameterized by a single complex parameter $\theta_\sigma$ and the Usadel equation \cite{Usadel} is:
\begin{align}
\frac{D_N}{2}(\partial_x^2+\partial_y^2)\theta_\sigma + iE\sin\theta_\sigma = 0,\label{eqUsadel} \\
\frac{D_F}{2}(\partial_x^2+\partial_y^2)\theta_\sigma + i(E+\sigma h)\sin\theta_\sigma = 0\label{eqUsadelF}
\end{align}
in N and F, respectively. In what follows we assume $\sigma=1$ and omit the spin index. Results for the opposite spin are then obtained by changing sign of $h$. 

Eqs.~\eqref{eqUsadel},\eqref{eqUsadelF} are supplemented with boundary conditions. At outer edges of the layers (in contact with vacuum or an insulating substrate) the condition is simply $\partial_y\theta=0$. The interface between layers is assumed transparent. This implies continuity of $\theta$ as well as conservation of current,
\begin{equation}
\nu_ND_N\partial_y\theta(x,d_N-0)=\nu_FD_F\partial_y\theta(x,d_N+0)\label{eqbcGeneral}
\end{equation}
where $\nu_i$ denotes the density of states in layer $i$. 

Our key assumption is that the layers are thin compared to all other length scales except mean free path. This means 
\begin{equation}
d^2_N\ll \frac{D_N}{E},\qquad d^2_F\ll \frac{D_F}{E+h}.\label{mainlimit}
\end{equation}
Here the typical energy $E$ is of the order of $E_\mathrm{Th}\equiv D/L^2$. Based on condition Eq.~\eqref{mainlimit} we expect the solution $\theta(x,y)$ to change relatively slowly with $x$ -- on length scales much larger than $d_N,d_F$. We also expect $\theta(x,y)$ to change weakly across the layers so that $\theta(x,y)$ has the form $\theta(x,y)=\vartheta(x)+\eta(x,y)$ with $\eta\ll\vartheta$. 

\textit{3.\ Effective boundary condition for thin F layer ---}
Consider the F layer at $d_N<y<d_N+d_F$. For simplicity we assume relatively strong exchange field, $|E|\ll |h|$. The function $\theta(x,y)$ in F can be approximated by 
\begin{equation}
\theta(x,y)\approx\vartheta_F(x)+\eta_F(x)\frac{(d_F+d_N-y)^2}{d_F^2}\label{eqAnsatzF}
\end{equation}
Substituting this into Eq.~\eqref{eqUsadelF} we get in the main order
\begin{align}
\frac{D_F\eta_F}{d_F^2}+ih\sin\vartheta_F=0.\label{eqEtaF}
\end{align}
All other terms were neglected assuming $\eta_F\ll \vartheta_F$ and $\partial_x^2\vartheta\ll h/D_F$. Eq.~\eqref{eqEtaF} expresses $\eta_F(x)$ through $\vartheta_F(x)$ and confirms that $\eta_F\ll\vartheta_F$ Eq.~\eqref{mainlimit}. We can now calculate $\partial_y\theta$ at the NF interface. Indeed, form Eq.~\eqref{eqAnsatzF} we have $\partial_y\theta\approx -2\eta_F/d_F=2ihd_F\sin\vartheta_F/D_F$. Substituting this into the general b.c. Eq.~\eqref{eqbcGeneral} and using $\theta(x,d_N-0)=\theta(x,d_N+0)\approx\vartheta_F(x)$ we obtain 
\begin{align}
\left.\partial_y\theta(x,y)\right|_{y=d_N-0}=2iq\sin\theta(x,d_N),\label{bcF}\\
q=\frac{hd_F\nu_F}{\nu_ND_N}.
\end{align}
We have thus reduced the F layer to an effective b.c.\ on $\theta$ in N. 

\textit{4. Effective equation for NF bilayer. --- }
We are now ready to study the solution of Eq.~\eqref{eqUsadel} with b.c.\ Eq.~\eqref{bcF} and $\partial_y\theta(x,y)|_{y=0}=0$. There are further b.c.\ at the NS interfaces at $x=0,L$, however, they are not important at the moment. It is sufficient to know that $L\gg d_N,d_F$ so that the typical length scale in the $x$ direction is large. 

We seek for a solution in the form 
\begin{equation}
\theta(x,y)=\vartheta(x)+\sum\limits_{n=2}^\infty \eta_n(x)\frac{y^n}{d_N^n}\label{eqAnsatzN}
\end{equation}
This ansatz formally incorporates all functions with $\partial_y\theta|_{y=0}=0$. Substituting this ansatz into Eq.~\eqref{eqUsadel} we get 
\begin{multline}
\vartheta(x)''+\sum\limits_{n=2}^\infty \left[\eta_n(x)''\frac{y^n}{d_N^n}+n(n-1)\eta_n(x)\frac{y^{n-2}}{d_N^n}\right]+\\\frac{2iE}{D_N}\sin\left(\vartheta(x)+\sum\limits_{n=2}^\infty \eta_n(x)\frac{y^n}{d^n_N}\right)=0\label{eqUsadelDetailed}
\end{multline}
where $\vartheta''\equiv\partial_x^2\vartheta$.
We can expand this two-dimensional equation in powers of $y/d_N$ and obtain a series of one-dimensional equations. For the first three powers, $y^0,y^1,y^2$ we get
\begin{align}
\vartheta''+\frac{2\eta_2}{d_N^2}+\frac{2iE}{D_N}\sin\vartheta=0\label{eqUNF0},\\
\eta_3=0,\label{eqUNF1}\\
\eta_2''+\frac{12\eta_4}{d_N^2}+\frac{2iE}{D_N}\eta_2\cos\vartheta=0,\label{eqUNF2}
\end{align}
respectively. The series can easily be continued. Odd powers, such as Eq.~\eqref{eqUNF1} lead to $\eta_{2k+1}=0$. Even powers, starting with Eq.~\eqref{eqUNF2} show that $\eta_{2k+2}\sim (d^2_NE/D_N)\eta_{2k}$ so that $\eta_{2k}$ form an exponentially decaying series.

The above equations Eqs. (\ref{eqUNF0}-\ref{eqUNF2}) are supplemented by the b.c. Eq.~\eqref{bcF}. Substituting our ansatz into it we get
\begin{align}
2\eta_2+4\eta_4=2iqd_N \left[\sin\vartheta+\eta_2\cos\vartheta\right]\label{bcNF}
\end{align}
where we only kept terms in the main two orders. We will solve Eqs.~\eqref{eqUNF0}-\eqref{bcNF} iteratively, starting with the main order. Neglecting the second term on both sides of Eq.~\eqref{bcNF} we get the main order solution which we denote $\overline{\eta}_2$:
\begin{equation}
\overline{\eta}_2=iqd_N\sin\vartheta.\label{eta2}
\end{equation}
Substituting this into Eq.~\eqref{eqUNF0} we get
\begin{gather}
\vartheta''+\frac{2i(E+h_\mathrm{eff})}{D_N}\sin\vartheta=0,\label{eqUmain}\\
h_\mathrm{eff}\equiv\frac{qD_N}{d_N}=h\frac{\nu_Fd_F}{\nu_Nd_N}.\label{heff}
\end{gather}
We see that in the main order $\vartheta(x)$ is governed by the one-dimensional Usadel Eq.~\eqref{eqUmain} which contains an effective exchange field $h_\mathrm{eff}$ induced by the F layer. This is a known result \cite{Golikova2012}. Our goal is to go beyond the approximation of Eq.~\eqref{eqUmain} and look for new physical effects from next-order corrections. 

To derive corrections to Eq.~\eqref{eqUmain} we need to calculate $\eta_2$ more accurately. We denote the correction to solution Eq.~\eqref{eta2} as $\delta\eta_2=\eta_2-\overline\eta_2$. From Eq.~\eqref{bcNF} we get (keeping only terms of the lowest non-vanishing order)
\begin{align}
\delta\eta_2=-2\eta_4+iqd_N\overline\eta_2\cos\vartheta.\label{deltaeta2}
\end{align}
$\eta_4$ can be expressed through $\eta_2$ using Eq.~\eqref{eqUNF2}:
\begin{align}
\eta_4=-\frac{d^2_N}{12}\left(\partial_x^2+\frac{2iE}{D_N}\cos\vartheta\right)\overline\eta_2=\\
-\frac{iqd^3_N}{12}\left(\vartheta''\cos\vartheta-(\vartheta')^2\sin\vartheta+\frac{iE}{D_N}\sin2\vartheta\right)=\label{line2}\\
-\frac{iqd^3_N}{12}\left(-\frac{i(2E+3h_\mathrm{eff})}{D_N}\sin2\vartheta-c\sin\vartheta\right).\label{line3}
\end{align}
Going from Eq.~\eqref{line2} to Eq.~\eqref{line3} we made use of Eq.~\eqref{eqUmain} to transform $\vartheta''$ and used the integral of motion $(\vartheta')^2-4i(E+h_\mathrm{eff})/D_N\cos\vartheta=\mathrm{const}=c$ to transform $(\vartheta')^2$. The term $c\sin\vartheta$ generates a small shift to the energy term in the Usadel equation Eq.~\eqref{eqUmain} and is therefore irrelevant. Omitting this term and substituting Eq.~\eqref{line3} into Eq.~\eqref{deltaeta2} we get
\begin{multline}
\delta\eta_2=\frac{qd^3_N(2E+3h_\mathrm{eff})}{6D_N}\sin2\vartheta-\frac{q^2d_N^2}2\sin2\vartheta\\=
\frac{d^4_NEh_\mathrm{eff}}{3D_N^2}\sin2\vartheta.
\end{multline}
Finally, substituting $\eta_2=\overline{\eta}_2+\delta\eta_2$ into Eq.~\eqref{eqUNF0} we obtain our final equation
\begin{gather}
\frac{D_N}{2}\vartheta''+i(E+h_\mathrm{eff})\sin\vartheta-\Gamma\sin2\vartheta=0,\label{eqFinal}\\
\Gamma=-\frac{d_N^2Eh_\mathrm{eff}}{3D_N}.\label{gamma}
\end{gather}
Equations \eqref{eqFinal} and \eqref{gamma} are the central result of our work. They show that superconductivity in a thin diffusive FN bilayer link is governed by an effective one-dimensional Usadel equation with an exchange field $h_\mathrm{eff}$ and a depairing term $-\Gamma\sin2\vartheta$. 

Equation \eqref{eqFinal} has to be supplemented with some boundary conditions at $x=0,L$. For example, for the setup Fig.~\ref{figSystem} with transparent interfaces and strong superconductors (e.g. thick leads to suppress the inverse proximity effect) the b.c.\ are $\vartheta(0)=\vartheta(L)=\pi/2$.

The magnitude of $\Gamma$ is small due to the $d_N^2/D_N$ factor. Thus, the depairing term is a negligible correction in most situations. 

Eq.~\eqref{eqFinal} governs the profile of $\vartheta(x)\equiv\theta(x,0)$. If we derive the same equation for $\theta(x,y_0)$ taken at a different $y_0$, for example $\theta(x,d_N)\equiv\vartheta+\eta_2+\eta_4+\dots$ we will arrive at a different coefficient $\Gamma'$. Moreover, Eq.~\eqref{gamma} implies that $\Gamma$ can have any sign, while a physical depairing term should have $\Gamma>0$ to suppress superconductivity. These observations indicate that the depairing term has no immediate physical meaning. However, it acquires clear physical meaning as soon as $E+h_\mathrm{eff}\ll |h_\mathrm{eff}|$ so that the depairing term becomes significant. In this case $\Gamma$ becomes observable in the sense that it governs the minigap $E_g$ of the S(N/F)S junction, as discussed in detail in the next paragraph. In this limit we have 
\begin{gather}
\Gamma=\frac{d_N^2h_\mathrm{eff}^2}{3D_N}.\label{gammasimple}
\end{gather}
which is positive, as required. Moreover, in this limit $\Gamma$ is the same for effective Usadel equations on $\theta(x,y_0)$ at all $y_0$ as expected.  

\textit{5.\ Applicability conditions ---} In addition to the thinness requirement for each layer, Eq.~\eqref{mainlimit}, the general one-dimensional equation \eqref{eqFinal} and its derivation require one additional condition we have not mentioned earlier: 
\begin{equation}
\nu_Fd_F\ll\nu_Nd_N.\label{thincondition}
\end{equation} 
This condition is neccessary for the approximate b.c.\ Eq~\eqref{bcF} to be applicable. Indeed, we derived Eq~\eqref{bcF} from the approximate ansatz Eq.~\eqref{eqAnsatzF} which only contained a quadratic term $\eta_F$, but not the quartic (let's call it $\eta_{4F}$), whereas in N we kept both $\eta_2$ and $\eta_4$. From the general b.c.\ Eq.~\eqref{eqbcGeneral} we conclude that this approach is valid if $\nu_FD_F\eta_{4F}/d_F\ll \nu_ND_N\eta_4/d_N$ which translates into Eq.~\eqref{thincondition}. 

For good measure we have also explicitly solved the bilayer problem without requiring Eq.~\eqref{thincondition}: we kept $\eta_{4F}$ in the ferromagnet and used the general b.c.\ Eq.~\eqref{eqbcGeneral}. This complicates the calculation considerably but does not lead to any new physical effects. In particular, both coefficients $D_\mathrm{eff}$ and $h_\mathrm{eff}$ in the effective Usadel equation are averages of the corresponding coefficients in the layers weighed with $\nu d$ in agreement with \cite{Golikova2012} e.g. $D_\mathrm{eff}=(\nu_Nd_ND_N+\nu_Fd_FD_F)/(\nu_Nd_N+\nu_Fd_F)$ while the result for $\Gamma$ is very cumbersome in the general case.

\textit{6.\ Minigap suppression in S(N/F)S junction ---} The minigap is a phenomenon exhibited primarily by SNS junctions \cite{minigapGolubov}. Long SNS junctions with good NS interfaces have a minigap $E_{g}$ in their spectrum \cite{minigapSpivak,minigapIvanov}, so that the density of states (DoS) $\rho_0(E)$ is zero at $|E|<E_{g}.$ For transparent NS interfaces $E_{g0}=C_2E_\mathrm{Th}$ where $E_\mathrm{Th}\equiv D/L^2$ is the Thouless energy and $C_2\approx3.122$.\cite{minigapSpivak,minigapIvanov}. In an SFS junction an exchange field $h$ is present which effectively shifts the spin-resolved DoS curve in energy by $\sigma h$ so that $\rho_\sigma(E)=\rho_0(E+\sigma h)$. However, the shape of the curve, and the width of the minigap in particular, remain intact. 

Adding depairing to an SNS junction affects its minigap. As $\Gamma$ increases the minigap width decreases and eventually vanishes at $\Gamma=\Gamma_c=\pi^2E_\mathrm{Th}/4$ \cite{minigapCrouzy}. At higher values only a dip in $\rho(E)$ remains in place of the minigap. This has been studied for SNS junctions with magnetic impurities \cite{minigapCrouzy,minigapHammer} which produce a depairing term identical to Eq. \eqref{eqFinal}. There is no macroscopic exchange field in this case so the minigap (or dip) is centered at $E=0$ at all times. 

In our S(N/F)S system we have both an exchange field and depairing. Therefore, the spin-resolved densities $\rho_\sigma(E)$ exhibit minigaps centered at $E_0=-\sigma h_\mathrm{eff}$ and these minigaps are shrunk or fully suppressed by $\Gamma$. 
For the suppression effect to be considerable $\Gamma$ must be of the same order as $E_\mathrm{Th}$. This means that $h_\mathrm{eff}$ must be of the order $\sqrt{E_\mathrm{Th}D_N}/d_N=D_N/(Ld_N)\gg E_\mathrm{Th}$. In particular, the critical value at which the minigap closes is
\begin{gather}
h_{c}=\frac{\sqrt{3D_N\Gamma_c}}{d_N}=\frac{\pi\sqrt{3} D_N}{2Ld_N}.\label{hc}
\end{gather}

\begin{figure}
\centering
\hspace*{-3pt}\includegraphics[width=0.485\textwidth]{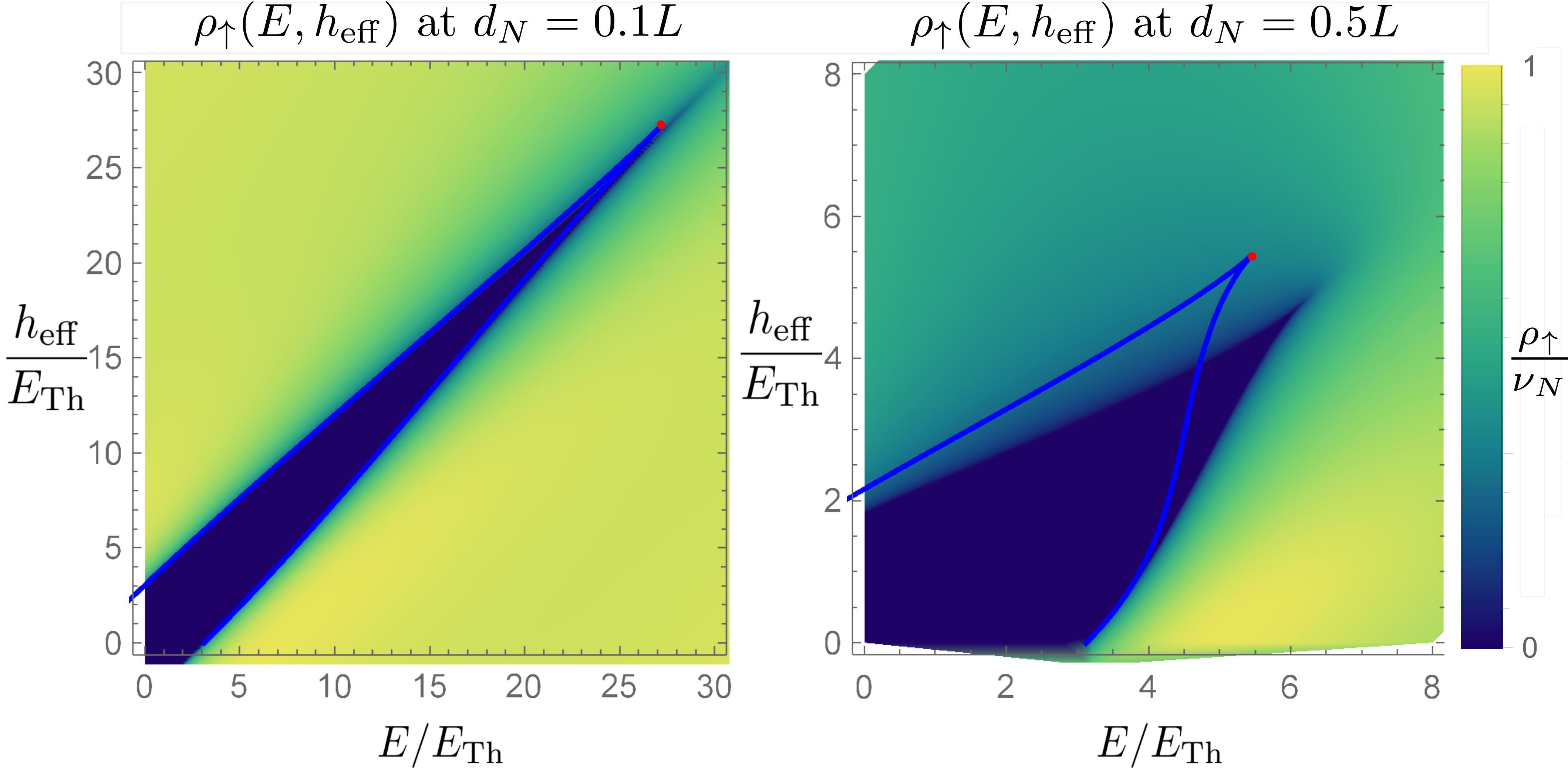}
\caption{DoS of spin-up states $\rho_\uparrow$ vs energy and effective magnetic field for $L=10d_N$ (left) and $L=2d_N$ (right). Solid blue lines show minigap calculated from our effective one-dimensional theory Eq.~\eqref{eqFinal}. The red dot indicates closing of the minigap at $h_\mathrm{eff}=h_c$. Color-coded data are from solving the two-dimensional Eq. \eqref{eqUsadel} directly. Dark blue means zero density of states, $\rho_\uparrow=0$ and yellow means normal metallic DoS, $\rho_\uparrow=\nu_N$. 
}
\label{figNumerics}
\end{figure}

Fig.~\ref{figNumerics} shows $\rho_\uparrow(E)$ in a S(N/F)S junction for different values of $h_\mathrm{eff}$. The DoS was calculated numerically by solving the original two-dimensional equations \eqref{eqUsadel},\eqref{eqUsadelF}. The solid line shows the minigap obtained from the effective 1D theory of \eqref{eqFinal}. As $h_\mathrm{eff}$ is increased the minigap shifts to the right and shrinks, eventually closing at $h_\mathrm{eff}=h_c$. For $L/d_N=10$ (left) the agreement between Eq.~\eqref{eqFinal} and two-dimensional numerics is very good, while for $L/d_N=2$ (right) it is also remarkable given that the small parameter of our theory is $d_N/L$.

\begin{figure}
\centering
\hspace*{-3pt}\includegraphics[width=0.485\textwidth]{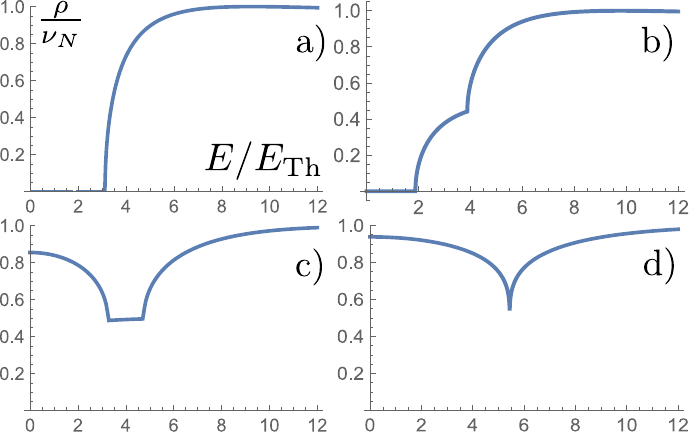}
\caption{DoS $\rho=\rho_\uparrow+\rho_\downarrow$ at $x=L/2$ of a S(N/F)S system with $L=2d_N$ for different values of $h_\mathrm{eff}$, illustrating the different regimes. $\rho$ is even with respect to energy: $\rho(-E)=\rho(E)$. The parameters are: a) $h_\mathrm{eff}=0$, so that $\rho=2\rho_\uparrow$ and there is a clear minigap b) $h_\mathrm{eff}=E_\mathrm{Th}<E_g$, the minigap edge is now staggered due to the Zeeman splitting c) $h_\mathrm{eff}=4E_\mathrm{Th}> E_g$, the minigaps of $\rho_\uparrow$ and $\rho_\downarrow$ are now well-separated, there are is no true gap in $\rho$ d) $h_\mathrm{eff}=h_c$. The minigap in the spin-resolved $\rho_\sigma$ has just closed and has become a dip. Further increase of $h_\mathrm{eff}$ will reduce the dip. }
\label{figGraphs}
\end{figure}

Using $\rho_\uparrow(E,h)=\rho_\uparrow(-E,-h)=\rho_\downarrow(-E,h)$ the total DoS can be expressed as
$\rho(E,h)=\rho_\uparrow(E,h)+\rho_\uparrow(-E,h).$ Thus $\rho(E,h)$ can be obtained by adding Fig.~\ref{figNumerics} to its mirror image. The resulting $\rho(E)$ is presented on Fig.~\ref{figGraphs}. Depending on the strength of the induced exchange $h_\mathrm{eff}$, the S(N/F)S system can be in one of four limits. At low field, $h_\mathrm{eff}\ll E_\mathrm{Th}$, the depairing is negligible and the only effect (compared to an SNS junction) is the spin-splitting of the mini-gap edge as seen on Fig.~\ref{figGraphs} a)b). At intermediate fields, $E_\mathrm{Th}\ll h\ll D/(d_NL)$, the mini-gaps in $\rho_\uparrow$ and $\rho_\downarrow$ are shifted considerably, leaving no overlap, so there is no true gap in the full spectrum, see Fig.~\ref{figGraphs}c). Finally, at large fields the mini-gaps shrink due to depairing and close at $h_c=\pi\sqrt{3}/2\cdot D/(d_NL)$, see Fig.~\ref{figGraphs}.d) At even higher fields there are no gaps in $\rho_{\uparrow},\rho_{\downarrow}$, only a dip at $E=\pm h$.

\textit{7. Discussion ---} Depairing terms as in Eq.~\eqref{eqFinal} are well-known. Typically, a depairing term occurs due to magnetic impurities \cite{AG}, i.e. random short-range exchange fields. It can also be produced by random exchange fields with longe-range fluctuations, (random magnetization domains)\cite{sfIvanovFominov,sfCrouzy,sfOstrovsky}. Orbital effects of magnetic fields can also result in this term \cite{Larkin65,BelzigDepairing}. Another source of a depairing term is mesoscopic fluctutaions of interaction strength/order parameter \cite{Feigel}. The common property of all these mechanisms is the averaging over different mesoscopic realizations to produce a description for the average system. In contrast, our S(N/F)S system has a fixed configuration of exchange fields. Instead of ensemble-averaging, our depairing term can be interpreted as the result of averaging over the $y$ coordinate, reducing the two-dimensional description to a one-dimensional equation on $\vartheta(x)$. 

The particular S(N/F)S setup Fig.~\ref{figSystem} we considered was motivated by experiment \cite{Golikova2012}. The experimental system had a weak link length $L=130\mathrm{nm}$ with layer thicknesses $d_N=60\mathrm{nm}$ (copper) $d_F=10-15\mathrm{nm}$ (iron). This puts it in the regime where our theory is applicable. The system is not particularly long, $L\approx2d_N$. However, even for this modest ratio our one-dimensional theory is in decent agreement with the solution of the original two-dimensional equation, see Fig.~\ref{figNumerics} (right).

Judging by the data of Ref~\cite{Golikova2012} and Fig.~4 therein, the sample is in the low-$h_\mathrm{eff}$ regime (top curve on Fig.~\ref{figGraphs}): a minigap of $E_g\approx 65\mu e\mathrm{V}$ is seen with its edge split by $\Delta U\approx 20\mu e\mathrm{V}$. Thus, depairing effects are negligible in this experiment.

For minigap suppression to be observed in the S(N/F)S junction, the order parameter $\Delta$ in S must be higher than $h_c$, otherwise the minigap would be pushed outside the superconducting gap of the leads. While this is achievable, there is an easier way. If we make an S(F/N/F')S junction where F and F' are similar layers with opposite magnetization, $h_\mathrm{eff}$ would vanish (or be small if F and F' are not identical enough). We have checked that the depairing is still present in such a trilayer and of the order $d^2_Nh^2_\mathrm{eff0}/D_N$ where $h_\mathrm{eff0}$ is the effective exchange field that would appear if F and F' had parallel magnetization. The absence of $h_\mathrm{eff}$ would also mean that the minigaps for different spins overlap, forming a true gap in the spectrum as on Fig.~\ref{figGraphs}a,b). Deatiled calculations for the S(F/N/F)S minigap will be published elsewhere. 

The effective Usadel equation Eq.~\eqref{eqFinal} is not limited to Josephson junction geometry, it can also be used for $S(N/F)$, $S(N/F)NS$ and other quasi-one-dimensional problems where one of the links is a thin bilayer oriented in parallel to the system axis. 

Apart from FN bilayers the theory also applies to N$\mathrm{I_F}$ bilayers where $\mathrm{I_F}$ stands for a magnetic insulator -- an insulator that reflects electrons with different spins differently. As shown by Cottet \textit{et al} \cite{Nazarov} the b.c.\ for the Usadel equation at such an interface has exactly the form of Eq.~\eqref{bcF}. 

Formally, Eq.~\eqref{bcF} can even be interpreted as a Kupriaynov-Lukichev b.c.\ \cite{KuLu} describing a tunnel contact with a normal metal with an imaginary tunnel conductance $g_t\propto iq$. Conversely, if we indeed consider an SNS junction where N is tunnel-coupled to a metallic reservoir, we should use Eq.~\eqref{bcF} with an imaginary $q$. This would produce an imaginary addition $i\gamma$ to energy (independent of spin) representing leakage into the metallic reservoir. In this case the minigap would disappear and the depairing term would be inconsequential since $(E+i\gamma)\sin\vartheta$ would always remain the dominant term in the effective Usadel equation. 

So far we have only discussed the effect of depairing on the DoS. Other key properties of Josephson junctions are their current-phase $I(\varphi)$ and voltage-current $V(I)$ characteristics. To calculate supercurrent one must solve the Usadel equation at non-zero phase difference $\varphi\neq0$, which adds an extra variable $\chi(x,y)$ which is the superconducting phase of the anomalous Green's function. We believe that in this case the two-dimensional Usadel equations can still be reduced to an effective one-dimensional theory, however this is a much more cumbersome problem since the number of variables is doubled. Such a calculation is beyond the scope of this letter. Nevertheless it is certain that the current-phase relation $I(\varphi)$ will have the same time-reversal symmetry $I(-\varphi)=-I(\varphi)$ as an SFS (and SNS) junction. This directly follows from symmetry of the S(N/F)S junction under $\{x,\varphi\}\mapsto\{-x,-\varphi\}$. We expect the depairing to only have a small quantitative effect on $I(\varphi)$ with $\delta I\sim I d^2_Nh_\mathrm{eff}/D_N$. This is supported by the following argument: the equilibrium supercurrent $I(\varphi)$ can be calculated via the Matsubara representation where $E=i\omega_k$ is purely imaginary. In this case the term $i\omega_k+h_\mathrm{eff}$ can never be small compared to $h_\mathrm{eff}$ and therefore the energy term in Eq.~\eqref{eqFinal} will always dominate the depairing term.

The voltage-current charasteristic $V(I)$ of Josephson junctions is typically quite rich and complicated. It usually reflects any peculiarities in the DoS, leading to features at $eV=2\Delta$ and $eV=2E_g$, as seen on Fig.~4 of Ref.~\cite{Golikova2012}, but can also exhibit additional phenomena such as subharmonic gap structure at $eV=2\Delta/n$ due to multiple Andreev reflection \cite{MARexp,MARtheory1,MARtheory2}. We do not know how multiple Andreev reflection, exchange field, depairing (and underlying bilayer inhomogeneity) would combine but we are confident that the minigap suppression we discussed would be reflected in the $V(I)$ of an S(N/F)S junction.

\textit{7. Conclusions ---}
In conclusion, we have shown that superconductivity in a thin NF bilayer can be described by a single one-dimensional Usadel eq.~\eqref{eqFinal} with an effective exchange field $h_\mathrm{eff}$ Eq.~\eqref{heff} and a depairing rate $\Gamma$ eq.~\eqref{gamma}. The depairing stems from the inherent bilayer inhomogeneity and suppresses superconductivity. The depairing rate is of the order $\Gamma\sim d_N^2h_\mathrm{eff}^2/D_N\ll h_\mathrm{eff}$ and only becomes important at energies close to $\pm h_\mathrm{eff}$. There it leads to a reduction of the minigap in the spin-resolved spectrum of the S(N/F)S junction. At $h_\mathrm{eff}>h_c=\pi\sqrt{3}D_N/(2Ld_N)$ the minigap closes completely. A promising system to observe the minigap effects would be an S(F/N/F')S junction where F and F' in the trilayer link have opposite magnetization. In this case the field $h_\mathrm{eff}$ can be cancelled out so that the spin-resolved minigaps coincide and are thus easier to observe. At the same time the depairing mechanism is still present in the FNF' trilayer leading to the suppression effects discussed. Apart from links made of N and F layers our theory can be applied to structures involving other materials, e.g.\ magnetic insulators.

\textit{8. Acknowledgements ---} We thank V.~V.~Ryazanov, Ya.~V.~Fominov, P.~M.~Ostrovsky and M.~V.~Feigel'man for valuable discussions. This work was supported by the Russian Science Foundation (Grant No. 19-72-00125). Numerical studies of the one-dimensional Usadel equation were supported by the Basic research program of Higher School of Economics.

\end{document}